\documentclass[a4paper]{rspublic}

\usepackage{siunitx}
\usepackage{url}
\usepackage{mathtools}
\usepackage{subfigure}
\usepackage[percent]{overpic}

\renewcommand{\vec}[1]{\ensuremath{\mathbf{#1}}}
\newcommand{\unitvec}[1]{\vec{\hat{#1}}}

\newcommand{\mat}[1]{\textrm{#1}}

\newcommand{\HemeLB}{\textsc{HemeLB}}

\newcommand{\trac}{\vec{t}}
\newcommand{\norm}{\unitvec{n}}

\usepackage{color}
\newcommand{\review}[1]{#1}

\begin{document}

% \title[Blood rheology and vascular risk]{Choice of blood rheology
%   model has minor impact on computational assessment of shear stress
%   mediated vascular risk} 

% \title[Blood rheology and vascular risk]{Choice of blood rheology
%   model has minor impact on the assessment of shear stress mediated
%   vascular risk \review{in a computational model of the middle
%     cerebral artery}}

\title[Blood rheology and vascular risk]{Impact of blood rheology on wall shear stress in a model of the middle cerebral artery}

\author[Bernabeu, Nash, Groen, Carver, Hetherington, Kr\"uger, Coveney]{Miguel
  O. Bernabeu$^{1,2}$, Rupert W. Nash$^{1}$, Derek Groen$^{1}$, Hywel
  B. Carver$^{1,2}$, James Hetherington$^{3}$, Timm Kr\"uger$^{1}$, Peter V. Coveney$^{1}$}

\affiliation{$^{1}$Centre for Computational Science, Department of
  Chemistry, University College London, 20 Gordon Street, London, WC1H
  0AJ, UK\\
$^{2}$CoMPLEX, University College London, Physics Building, Gower Street,
London, WC1E 6BT, UK\\
$^{3}$Research Software Development Team, Research Computing and
Facilitating Services, University College London, Podium Building -
1st Floor, Gower Street, London, WC1E 6BT, UK}
%\email[Corresponding author: ]{miguel.bernabeu@ucl.ac.uk}

\label{firstpage}

\maketitle

\begin{abstract}{Blood flow modelling, rheology,
    multiscale modelling, lattice-Boltzmann, three-band diagram
    analysis}
  Perturbations to the homeostatic distribution of mechanical forces
  exerted by blood on the endothelial layer have been correlated with
  vascular pathologies including intracranial aneurysms and
  atherosclerosis. Recent computational work suggests that in order to
  correctly characterise such forces, the shear-thinning properties of
  blood must be taken into account. To the best of our knowledge,
  these findings have never been compared against experimentally
  observed pathological thresholds. In the current work, we apply the
  three-band diagram (TBD) analysis due to Gizzi \emph{et al.} to
  assess the impact of the choice of blood rheology model on a
  computational model of the right middle cerebral artery.  Our
  results show that\review{, in the model under study,} the differences between the wall shear stress
  predicted by a Newtonian model and the well known Carreau-Yasuda
  generalized Newtonian model are only significant if the vascular
  pathology under study is associated with a pathological threshold in
  the range \SIrange{0.94}{1.56}{\pascal}, where the results of the
  TBD analysis of the rheology models considered differs. Otherwise,
  we observe no significant differences.

\end{abstract}

%\todo{Articles are to be no more than 6,000 words in length (including references) and figures should be able to occupy a single side of A4.}
\section{Introduction}

Physiology and medicine are being revolutionized by the growing role
of information technology. Our ability to acquire and manage data on
both animals and humans allows us to develop increasingly detailed
computational models of the biological processes sustaining
life. These models, together with the relevant experimental data, are
helping researchers to gain insight into the physiology and pathology
of the systems under study, in many cases beyond what is possible with
purely observational methods.

Cerebrovascular disorders, including intracranial aneurysms (ICA),
which may rupture leading to subarachnoid haemorrhages, are one of the
most prevalent and devastating diseases of adults, of worldwide
concern. In the UK, the total burden has been estimated as
\pounds{}0.5 billion annually \cite{rivero-arias10}. It is currently
generally accepted that haemodynamics plays an important role in the
appearance, evolution, and potential rupture of these type of vascular
pathologies \cite{stehbens89, shojima04}. More precisely, perturbations in the
homeostatic distribution of mechanical forces exerted by the blood
on the endothelial layer have been
correlated not only to aneurysm initiation and rupture but also to the
development of other vascular pathologies such as atherosclerosis
\cite{chatzizisis08}.

From a rheological point of view, blood is a shear-thinning fluid.
%which flows only when subject to stresses greater than a certain
%yield stress. 
This behaviour arises from the presence of red blood cells suspended
in a medium known as blood plasma. Despite this
fact, a large body of literature concerning computational
haemodynamics characterises blood as a Newtonian fluid under the
assumption that in large arteries the shear rate is large enough for
the viscosity to be treated as effectively constant \cite{mejia11}.

There has been increasing interest during recent years in the
comparison of blood rheology models using computational fluid dynamics
(CFD) simulations in realistic computational domains reconstructed
from medical image. These studies have been performed in both healthy
vasculature (see \emph{e.g.} \cite{morbiducci11, box05, johnston04})
and in the presence of aneurysms (see \emph{e.g.} \cite{bernsdorf09,
  cavazzuti11, fisher09}). In particular, authors pay special
attention to the influence of the choice of rheology model on the
computational estimates of wall shear stress (WSS) as a proxy for
aneurysm rupture risk.  It has been suggested that, in ICAs, the
Newtonian simplification overestimates WSS (see \emph{e.g.}
\cite{bernsdorf09, xiang12}) and may underestimate rupture
risk. Furthermore, it has been argued that applications requiring
accurate WSS estimates (\emph{e.g.}  those concerning vascular
remodelling and biomechanics) will suffer from modelling inaccuracy
unless the generalized Newtonian (GN) properties of blood are taken
into account.

Different indices have been proposed as a way of integrating into
CFD-based biomarkers of rupture risk the rich spatio-temporal
structure of WSS induced by pulsatile flow in complex vascular
networks. Minimal or maximal peak WSS, time-averaged WSS, and
oscillatory shear index (OSI) \cite{ku85} have been extensively used,
among others. In many of the studies cited previously, comparisons are
performed based on colour map plots, of one or more of these indices,
on the surface of the original computational domain. Differences
between WSS estimates produced by different rheology models are hence
presented in a very qualitative way. Furthermore, the link to actual
rupture risk is based on the currently accepted low shear stress
biomarker, but with no actual correlation to experimentally determined
WSS thresholds. Several authors \cite{kallmes12, cebral12} have
recently criticised this approach to rupture risk quantification
arguing that more quantitative methods are required in order to gain
further insight into the problem.

In a recent publication, Gizzi \emph{et al.} \cite{gizzi11} proposed a
new framework for the quantitative analysis of WSS: the so-called
three-band diagram (TBD) analysis. The TBD analysis facilitates the
evaluation of the WSS obtained at a given location over time
(\emph{i.e.} a WSS signal) against a range of WSS pathological
thresholds (\emph{e.g.} WSS magnitude lower than \SI{0.5}{\pascal} as
reported in the case of atherosclerosis formation
\cite{chatzizisis08}). The analysis determines how likely a given WSS
signal is to be considered risky for any given threshold.
Furthermore, the results of the analysis can be easily compared
against pathological values of WSS observed experimentally.

In the current work, we apply TBD analysis to assess the impact of the
choice of blood rheology model on the WSS estimates of the
\HemeLB{}~\cite{mazzeo08} lattice-Boltzmann blood flow solver in a
high resolution three-dimensional model of the right middle cerebral
artery. The rest of the paper is structured as follows: \S2 introduces
the computational and mathematical models used in this work as well as
the simulation workflow implemented, \S3 presents the results of our
simulation and their main implications, finally \S4
summarises the main conclusions of the work and outlines future research directions.

% Our results show that the differences between the WSS signals
% predicted by a Newtonian blood rheology model and the Carreau-Yasuda
% model are only significant if the vascular pathology under study is
% associated with a risk factor in the range
% \SIrange{0.94}{1.56}{\pascal}, where the TBD analysis of the rheology
% models considered differs. Otherwise, no significant differences are
% identified and one can therefore use the computationally less expensive
% Newtonian model for simulations of blood flow in such geometries.

\section{Methods}

\subsection{3D model of the middle cerebral artery}
\subsubsection{Geometry generation}
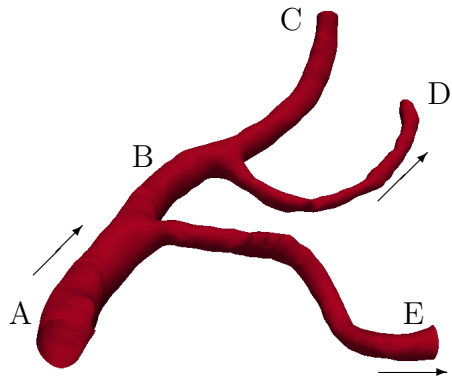
\begin{figure}
  \centering
  \begin{overpic}[scale=0.25]{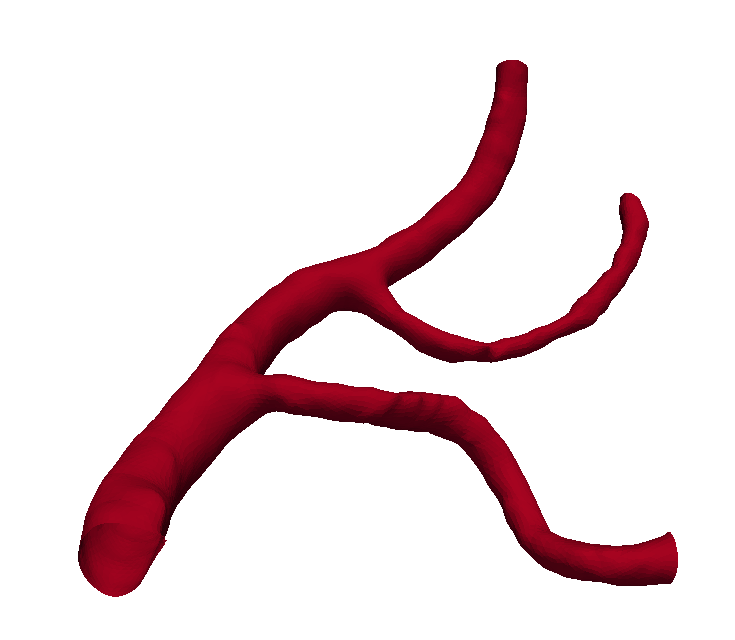}
    \put(5,15){\large A}
    \put(30,47){\large B}
    \put(60,75){\large C}
    \put(90,60){\large D}
    \put(85,15){\large E}
    \put(10,25){\vector(1,1){10}}
    \put(80,5){\vector(1,0){14}}
    \put(80,40){\vector(1,1){10}}
  \end{overpic}
  \caption{3D model of a subset of the right middle cerebral artery
    used in this work. Geometry segments are labelled A--E for later
    reference. The arrows indicate flow direction.}
  \label{fig:3DMCA}
\end{figure}

The three-dimensional (3D) model of the middle cerebral artery used in
this work (see figure \ref{fig:3DMCA}) is a subset of a geometrical
model of the intracranial vasculature reconstructed from rotational
angiography scans. % of a patient suffering from an intracranial
                  % aneurysm. 
It corresponds to a section of the right middle
cerebral artery (MCA) in the vicinity of the internal carotid
artery. The main geometrical features of the model are: i) vessels of
variable diameter, ii) two bifurcations, and iii) vessel bending. The
National Hospital for Neurology and Neurosurgery, London (UK) provided
the original images in the framework of the GENIUS project
\cite{mazzeo10} as part of a larger dataset library. The dataset used
in this work was segmented and the surface mesh in figure
\ref{fig:3DMCA} generated with the open source package Vascular
Modeling Toolkit (VMTK) \cite{antiga08}.

\subsubsection{\HemeLB}
\HemeLB{}~\cite{mazzeo08} is an open source software
platform\footnote{The codebase is available under LGPL license from
  \url{http://ccs.chem.ucl.ac.uk/hemelb}.} for modelling and
simulation of blood flow in sparse vascular networks. It is comprised
of tools for geometrical model preprocessing (\emph{i.e.} regular grid
volume meshing of surface meshes), simulation on massively parallel
architectures, real-time visualization and steering, and data
post-processing. To date, \HemeLB{} has been successfully applied to
the simulation of blood flow in healthy brain vasculature as well as
in the presence of intracranial aneurysms. Particular attention has
been paid to obtaining and presenting simulation results in a
clinically meaningful way \cite{mazzeo10}. \HemeLB{} uses the lattice-Boltzmann method
for fluid dynamics (see \emph{e.g.} \cite{chen98}) since it allows
efficient implementations in large-scale high performance computing
infrastructures. For this work, we have developed an extension of
\HemeLB{}'s lattice Bhatnagar-Gross-Krook (LBGK) collision operator in
order to accommodate both Newtonian and GN rheology models. We use the
D3Q15 velocity set and the halfway bounce-back rule \cite{chen98} to
enforce the no-slip boundary condition at the walls. We have recently
shown~\cite{carver12a} that this combination of collision operator,
velocity set, and wall boundary condition performs well from both a
numerical and a computational point of view in complex domains for
typical blood flow Reynolds and Womersley numbers. \review{Our results 
show that first-order convergence of the velocity field is achieved over a
wide range of resolutions and Reynolds numbers.}

\subsubsection{Generalized Newtonian rheology}
The Carreau-Yasuda (CY) model is widely used to describe the
shear-thinning behaviour of blood \cite{boyd07,ashrafizaadeh09}. In
this model, the dynamic viscosity $\eta$ is related to the shear rate
$\dot{\gamma}$ through the following expression:
\begin{equation}
\eta(\dot{\gamma}) = \eta_{\infty} + ( \eta_0 - \eta_{\infty})\left( 1
  + (\lambda \dot{\gamma})^a \right)^{\frac{n-1}{a}}\;,
\label{eq:cy_model}
\end{equation}
where $a$, $n$, and $\lambda$ are empirically determined to fit a
curve between regions of constant $\eta_{\infty}$ and $\eta_0$. This
model defines three different regimes: a Newtonian region with
$\eta_0$ for low shear rate, followed  by a shear-thinning
region where $\eta$ decreases with $\dot{\gamma}$; finally, once
$\eta_{\infty}$ is reached a third Newtonian region with constant viscosity
$\eta_\infty$ is defined for high shear rates. In this work, we will use
the values provided in \cite{boyd07}: $\eta_0=0.16$ \si{\pascal\second},
$\eta_{\infty}=0.0035$ \si{\pascal\second}, $\lambda = 8.2$ \si{\second},
$a=0.64$, and $n=0.2128$ (both are dimensionless).

\begin{figure}[htbp]
  \begin{center}
    % For pdflatex compilation use:
    % gnuplot viscosity_dyn.gp && epstopdf viscosity_dyn-inc.eps && pdflatex viscosity_dyn
    % For standard latex use:
    % gnuplot viscosity_dyn.gp && latex viscosity_dyn
    \includegraphics[width=0.8\textwidth]{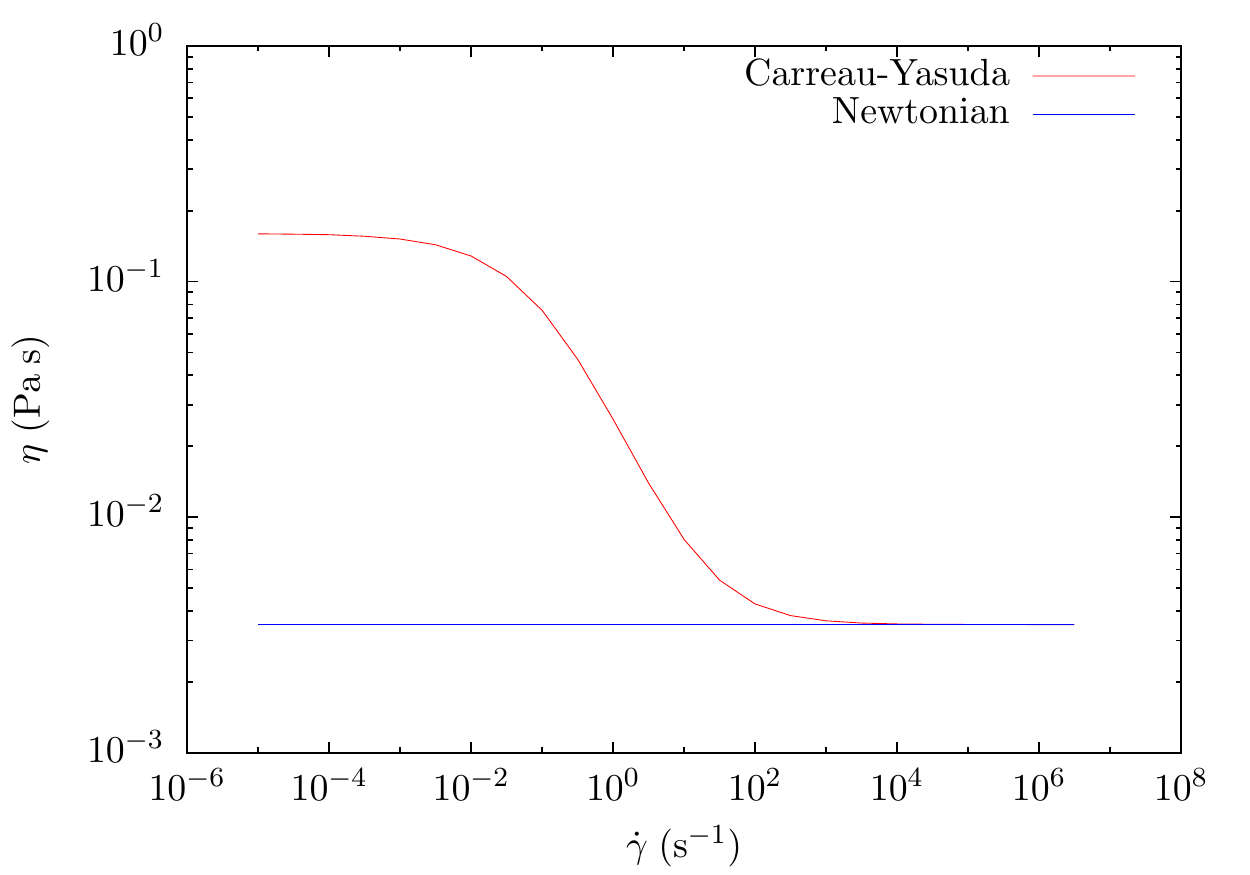}
    \caption{Dynamic viscosity $\eta$ as a function of shear rate
      $\dot{\gamma}$ for the Carreau-Yasuda (red) and the Newtonian
      (blue) models.}
    \label{fi:shear_viscosity}
  \end{center}
\end{figure}

Figure \ref{fi:shear_viscosity} presents
viscosity as a function of shear rate for the previous model and the
Newtonian model considered in this work
($\eta=\SI{3.5e-3}{\pascal\second}$). The Carreau-Yasuda model displays
a smooth transition between $\eta_0$ and
$\eta_{\infty}$. Significant haemodynamic differences between the two
rheology models are expected for simulations with $\dot{\gamma} <
\SI{100}{\per\second}$.

\subsection{1D model of the human vascular system}

In order to obtain inlet and outlet boundary conditions for
our 3D model, we use a modified version of a one-dimensional (1D) model of
the human vascular system previously published in \cite{mulder11}. The 
original model includes a 1D representation of the main
arteries in the upper body (including the circle of Willis and both
MCAs) and a zero-dimensional representation of the peripheral
vasculature. The geometry is available as part of the open source
software package pyNS \cite{manini12} which also implements a numerical solver for
the 1D pulse propagation mathematical model presented in
\cite{huberts12}.

\begin{figure}
  \centering
  \begin{overpic}[scale=0.25]{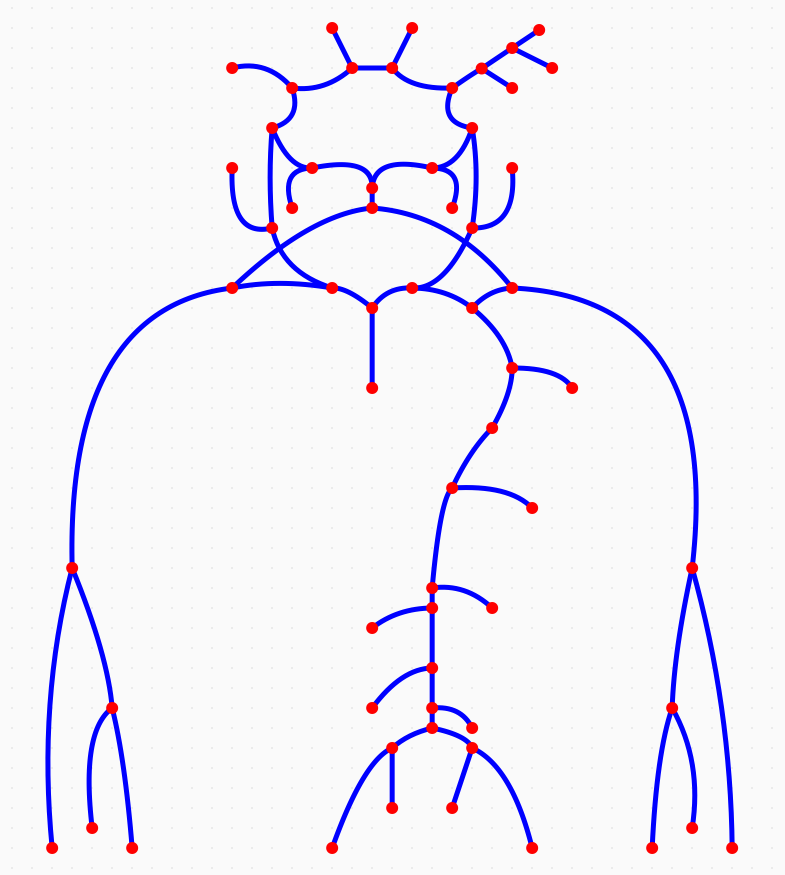}
    \put(40.25,50){\Large H}
    \put(75,95){\vector(-1,0){10}}
  \end{overpic}
  \caption{1D model of the main arteries in the upper body. Each
    segment (in blue) represents a different part of a human arterial
    system including characterisation of the 3D model of the right MCA
    in figure \ref{fig:3DMCA} (see top right arrow). Segment lengths
    are not to scale. The red dots represent: i) bifurcations, when
    found at the intersection of two or more segments, ii)
    zero-dimensional representations of the peripheral vasculature,
    when found at the end of an open-ended segment or iii) the heart
    in the case labelled with H.}
  \label{fig:1DModel}
\end{figure}

The original 1D geometrical model does not include any detail about
the vessels branching off the right MCA. Therefore, we modified it to
include the two MCA 
side branches present in the 3D model (see figure
\ref{fig:3DMCA}). In order to achieve this a 1D characterisation of
the 3D geometry was required. The open source software package VMTK
was used to compute: a) the length of the centrelines associated with
segments A--E in figure \ref{fig:3DMCA}, and b) the radii of the maximum inscribed
spheres along the centrelines. Table \ref{tab:length_radii} summarises
the results and figure \ref{fig:1DModel} shows a schematic
of the 1D model including the more detailed representation of the
right MCA.
The reader is referred to \cite{mulder11} or the pyNS tool for the names
and characteristics (\emph{e.g.} length, radius, connectivity) of each
of the segments of the vascular system. 

\begin{table}
  \centering
  \caption{Geometrical characterisation of the 3D model used in this work.}
  \label{tab:length_radii}
  \begin{tabular}{cccc}
    \hline
    Segment & Length & Proximal end radius & Distal end radius \\
    \hline
    A & \SI{8.9}{\milli\metre}  & \SI{1.30}{\milli\metre} & \SI{1.25}{\milli\metre} \\
    B & \SI{11.2}{\milli\metre}  & \SI{1.25}{\milli\metre} & \SI{1.10}{\milli\metre} \\
    C & \SI{13.6}{\milli\metre}  & \SI{1.10}{\milli\metre} & \SI{0.85}{\milli\metre} \\
    D & \SI{20.6}{\milli\metre}  & \SI{0.62}{\milli\metre} & \SI{0.62}{\milli\metre} \\
    E & \SI{21.1}{\milli\metre}  & \SI{0.77}{\milli\metre} & \SI{0.77}{\milli\metre} \\
    \hline    
  \end{tabular}
\end{table}

%\subsection{1D-3D model coupling}

\subsection{Three-band diagram analysis}

The extraction of synthetic biomarkers of aneurysm rupture risk based
on CFD analysis of patient-specific models is an active field of research. 
The three-band diagram analysis framework was recently proposed \cite{gizzi11}
as a way of generalising previously proposed indices
(\emph{e.g.} time-averaged WSS, OSI) and providing a quantitative way of
comparing temporal WSS signals against specific risk factors.

Let $\trac$ be the instantaneous traction vector at an arbitrary surface point with
associated surface unit normal $\norm$ such that
\begin{equation}
\trac_i = \mat{T}_{ij} \norm_j\;,
\end{equation}
where $\mat{T}$ is the deviatoric part of the full stress tensor of a
fluid computed from the mesoscopic LB simulation variables as
described in \cite{krueger09}. The relationship between $\mat{T}$ and $\sigma$,
the full stress tensor, is
\begin{equation}
\sigma_{i j} = -P\delta_{i j} + \mat{T}_{i j}\;,
\end{equation}
where $P$ is the hydrodynamic pressure and $\delta$ is the usual
Kronecker delta tensor. Furthermore, for a generalized Newtonian
incompressible fluid, $\mat{T}$ can be rewritten as
\begin{equation}
\mat{T}_{i j} = 2\eta(\dot{\gamma}) \mat{S}_{i j}\;,
\end{equation}
where $\mat{S}$ is the shear rate tensor
\begin{equation}
\mat{S}_{i j}=\frac{1}{2}(\partial_iv_j+\partial_jv_i)\;,
\end{equation}
$\eta$ is the dynamic viscosity of a fluid as a function of
shear rate $\dot{\gamma} = \sqrt{2 S_{ij}S_{ij}}$, see \emph{e.g.}
equation \eqref{eq:cy_model}, and $\vec{v}$ is the velocity vector. In the case of Newtonian fluids,
$\eta(\dot{\gamma}) = \eta = \const$

In this work, we will consider the temporal
evolution of the signed magnitude of the traction vector $\trac$
(which we will refer as the WSS signal), \emph{i.e.}:
\begin{equation}
S(t) \coloneqq \sgn({\trac(t) \cdot \overline{\trac}}) |\trac(t)|\;,
\end{equation}
where $\overline{\trac}$ is the average traction vector over time, $\sgn$ is
the sign function, and $|\cdot|$ is the magnitude of a given vector.

Given a WSS signal, $S(t)$, and a scalar risk factor $\sigma \geq 0$, the TBD
analysis defines a triplet of functions
\begin{align}
S^+(t)& \coloneqq S(t) H(S(t) - \sigma), \nonumber \\
S^0(t)& \coloneqq S(t) H(\sigma - S(t)) H(\sigma + S(t)), \nonumber \\
S^-(t)& \coloneqq S(t) H(-S(t) - \sigma),
\end{align}
where the Heaviside function $H(x)$ is defined as $H(x) = 1$ if $x >
0$ and $H(x) = 0$ if $x < 0$.
% where $H^+(\sigma)=1$ if $S>\sigma$ and 0 otherwise, $H^0(\sigma)=1$
% if $-\sigma\leq S\leq\sigma$ and 0 otherwise, and $H^-(\sigma)=1$ if
% $S<-\sigma$ and 0 otherwise. 
The closed support of the function $S^+$
is a set of time intervals of cardinality $N^{+}(\sigma)$ (and
similarly with $S^0$ and $S^-$). The main idea behind the method is
to inspect the number of such intervals as a function of the variable
threshold $\sigma$ (\emph{i.e.} the risk factor). It has been
suggested that a WSS signal can be considered healthy if
$N^{+,0,-}(\sigma)>0$ for a given risk factor $\sigma$. 
The reader is referred to \cite{gizzi11} for a more detailed description
of the method.
A Python implementation of the algorithm is freely available as part
of \HemeLB{}'s postprocessing tools.

\subsection{Simulation workflow}

Our simulation workflow is as follows. First, the original DICOM
images were segmented, a region of interest was chosen, and a surface
mesh was generated with the VMTK implementation of the marching cubes
algorithm. A non-shrinking Taubin filter was then applied to smooth
out imaging artifacts. Second, we loaded the resulting surface mesh in
\HemeLB{}'s setup tool in order to choose the location of the inlet
and the outlets and generate a regular grid discretisation of the
volume enclosed by the surface (with a total of 4,161,046 grid
points). Next, the pyNS solver was run to obtain pressure traces at
inlet A and oulets C--E in figure \review{\ref{fig:3DMCA}}. We then
used these traces as inlet and outlet boundary conditions to run
\HemeLB{} simulations with both the Newtonian and Carreau-Yasuda
rheology models for a total of three cardiac cycles. \review{Note that
  for flow in a domain with open boundaries, one must decide whether
  to close the system with pressure or flow rate boundary conditions. The
  former case ensures that the same pressure drop occurs regardless of
  the rheology model chosen but sacrifices the Reynolds number parity
  as a mathematical necessity. The latter would ensure the same
  Reynolds number is recovered but different pressure drops occur. We
  chose to impose pressure boundary conditions as obtained from the 1D
  simulations.}

The following
configuration paremeters were used in this work: timestep $\Delta
t=\SI{2.5706e-6}{\second}$, spacestep $\Delta x =
\SI{3.50e-5}{\metre}$, maximum density difference in the domain
$\Delta
\rho/\rho_0=0.019$, % This is the value in the spreadsheet, due to
                   % issues in the current iolet BC implementation it
                   % goes up to 0.031 in the simulation
Mach numbers (defined as the ratio between the largest velocity
magnitude in the domain and lattice speed of sound $c_s=\frac{\Delta
  x}{\sqrt{3}\Delta t}$) $0.31$ and $0.244$ for the Newtonian and CY
rheology models respectively, and LBGK relaxation parameter
$\tau=0.522$ for the Newtonian model and $\tau =
\frac{1}{2}+\frac{\eta(\dot{\gamma})}{\Delta t c_s^2 \rho}$ for the CY
model (which becomes $\tau=0.522$ for $\eta_{\infty}$). Finally, we
post-processed the simulation results to compute TBDs of the WSS
signal at different geometrical locations over time. Table
\ref{ta:sampling_points} shows their location for reproducibility
purposes.

\begin{table}
  \centering
  \caption{Three-band diagram analysis sampling points.}
  \label{ta:sampling_points}
  \begin{tabular}{cc}
    Point  & Coordinates \\ \hline
    $x_1$ & (11.38, 36.89, 45.17) \si{\milli\metre} \\
    $x_2$ & (10.98, 41.64, 47.97) \si{\milli\metre} \\
    $x_3$ & (12.77, 41.15, 45.02) \si{\milli\metre} \\
    $x_4$ & (12.34, 42.69, 46.17) \si{\milli\metre} \\
    $x_5$ & (11.88, 50.29, 41.31) \si{\milli\metre} \\
    $x_6$ & (11.59, 50.55, 49.19) \si{\milli\metre} \\
    $x_7$ & (12.89, 55.45, 44.75) \si{\milli\metre} \\
    $x_8$ & (12.82, 55.33, 45.58) \si{\milli\metre} \\
    $x_9$ & (14.57, 55.73, 52.63) \si{\milli\metre} \\
\hline
  \end{tabular}
\end{table}

The pyNS simulations were run on a single core of a 2 GHz Intel Core
i7 laptop with 4GB of RAM and took on the order of minutes to run. The
\HemeLB{} simulations were run on 2048 cores of HECToR (UK's national
supercomputer) and took \SI{53}{\minute} and \SI{83}{\minute} for the
Newtonian and Carreau-Yasuda models respectively. The difference in
computing time between the two rheology models is due to the cost
associated with computing $\dot{\gamma}$ and evaluating equation
\eqref{eq:cy_model} at each lattice site every timestep. The choice of
core count is based on the number of lattice sites and the scalability
analysis presented in \cite{groen12} in order to ensure efficient use
of the computational resources.

All the files required to run the pyNS and \HemeLB{} simulations are
available as part of the supplementary material associated with this
paper. Both software packages are open source and freely available to
the public.

\section{Results and discussion}

\subsection{Pressure profiles}
\begin{figure}
  \centering
  \includegraphics[scale=0.75]{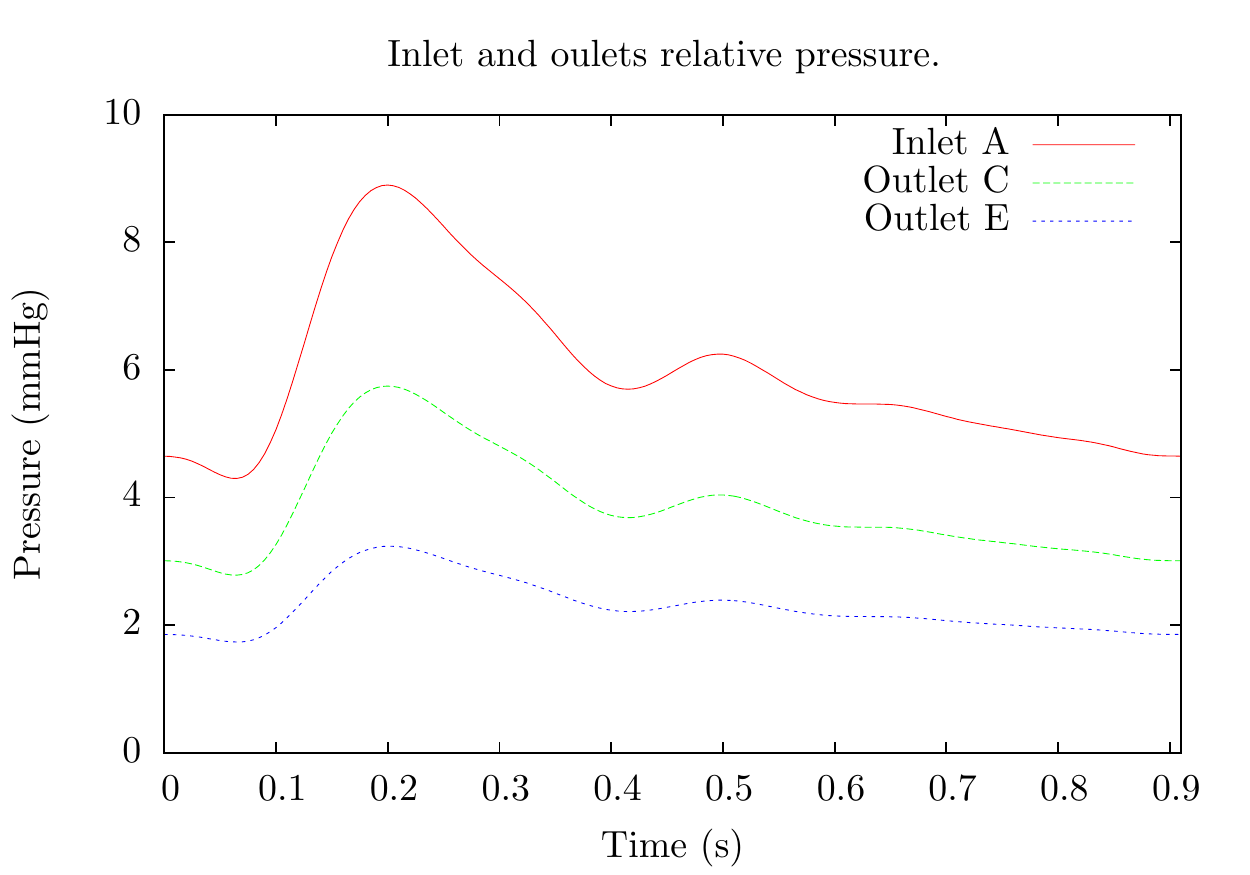}
  \caption{\review{Pressure differentials, relative to outlet D,
      obtained from the 1D model at the inlet and outlets.}
    Simulations were run for a total of three cardiac cycles.}
  \label{fig:pressure_profiles}
\end{figure}

Figure \ref{fig:pressure_profiles} plots the pressure traces generated
by pyNS at inlet A and outlets C--E of figure \ref{fig:1DModel}. It can be
seen how the typical pressure variations observed throughout the
cardiac cycle are well recovered. First, the sudden 
increase in pressure following the opening of the aortic valve after
approximately \SI{0.06}{\second}, which peaks at around
\SI{0.2}{\second} and drops until approximately
\SI{0.4}{\second} corresponds to the ventricular ejection
phase. Secondly, the isovolumic relaxation phase happens between the
previous time point and the mitral valve opening around
\SI{0.55}{\second}. Finally, the ventricular filling phase occurs
between the time the mitral valve opens and the end of the cycle at
approximately \SI{0.91}{\second} (for a cardiac rate of roughly
66 beats per minute). 

% Furthermore, figure \ref{} presents the pressure differences between
% the inlet/outlets A, C, and E and the outlet D, used as reference
% pressure in the simulations. It can be observed how the largest
% pressure differences in the domain occur after \SI{10000}{\second} of the
% beginning of each cardiac cycle 

\subsection{3D simulations and three-band diagram analysis}

\begin{figure}
  \centering
  \subfigure[End of diastole $t=\SI{1.82}{\second}$.]{
    \label{}
    \begin{overpic}[scale=0.25]{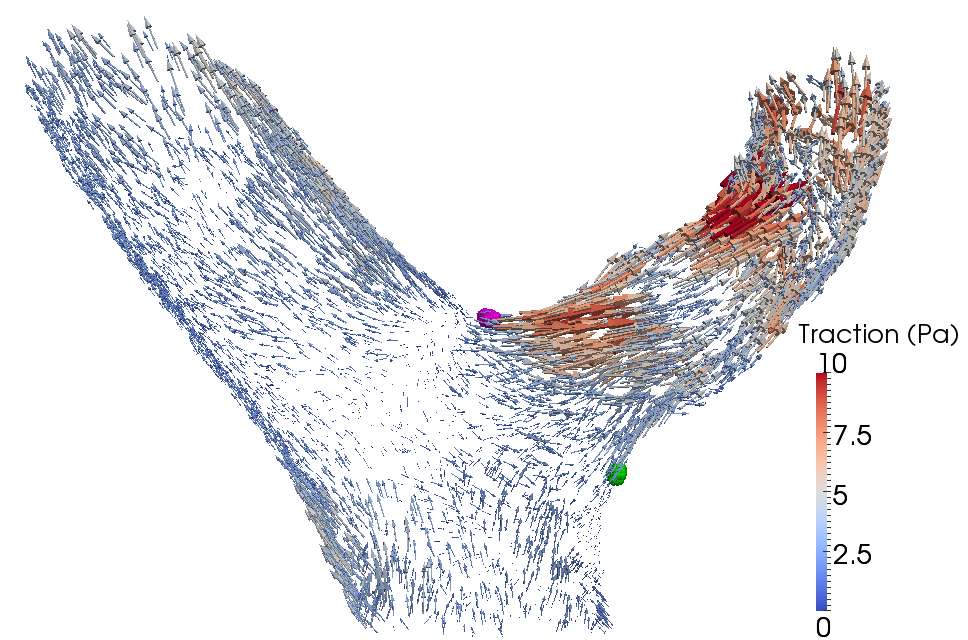}
      \put(49,36){$x_4$}
      \put(66,17){$x_3$}
    \end{overpic}
  }
  \subfigure[Peak flow during systole $t=\SI{1.97}{\second}$.]{
    \label{}
    \begin{overpic}[scale=0.25]{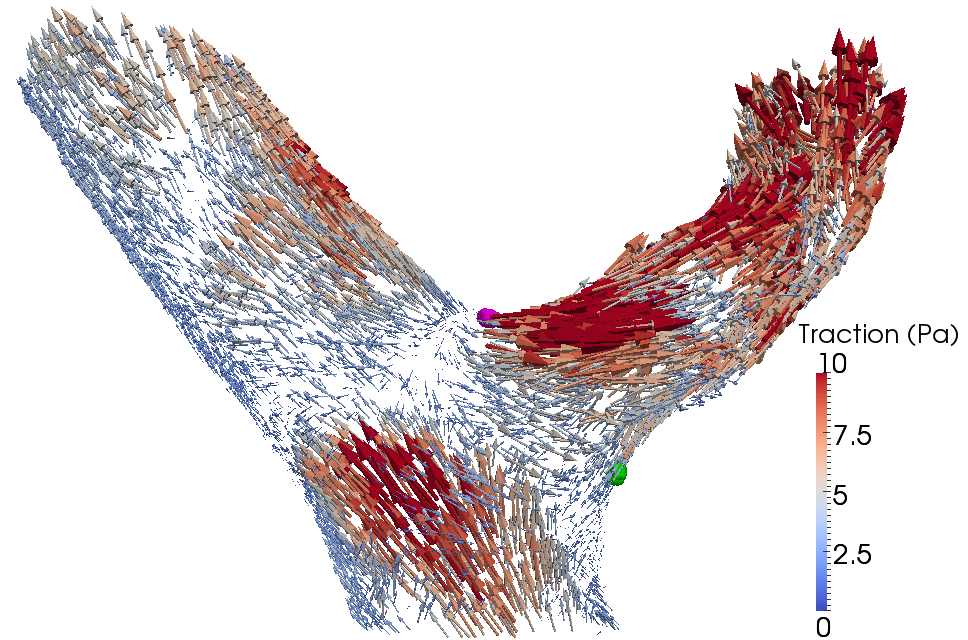}
      \put(49,36){$x_4$}
      \put(66,17){$x_3$}
    \end{overpic}
  }
  \caption{Traction vector $\trac$ (estimated with the CY rheology
    model) at the upstream bifurcation of the 3D model. Vectors are scaled
    according to their magnitude. Points $x_3$ and $x_4$ in table
    \ref{ta:sampling_points} are shown in green and magenta
    respectively. Visualizations generated with the open source
    software package  Paraview \cite{paraview}.} % A. Henderson, ParaView Guide, A Parallel Visualization Application. Kitware Inc., 2007.
  \label{fig:3D_traction}
\end{figure}

Figure \ref{fig:3D_traction} plots the traction vector $\trac$
computed with \HemeLB{} configured to use the CY rheology model at the
end of diastole and at peak flow in systole (both in the third cardiac
cycle simulated). The direction of $\trac$ (which is consistent with
the flow direction at any given point) and its magnitude follow a
complex distribution throughout the domain with notable changes during
the cardiac cycle. The main characteristics are: i) areas with larger
WSS can be found in the interior side of both branches leaving the
bifurcation, ii) a zone of low WSS, or stagnation point, is found
around the area where both branches meet, iii) there are substantial
changes in the direction of $\trac$ throughout the cardiac cycle
around the stagnation point area, leading to oscillatory WSS signals.

\begin{figure}
  \centering
  \subfigure[Newtonian TBD at $x_3$.]{
    \label{fig:tbd_n3}
    \includegraphics[width=0.4\textwidth]{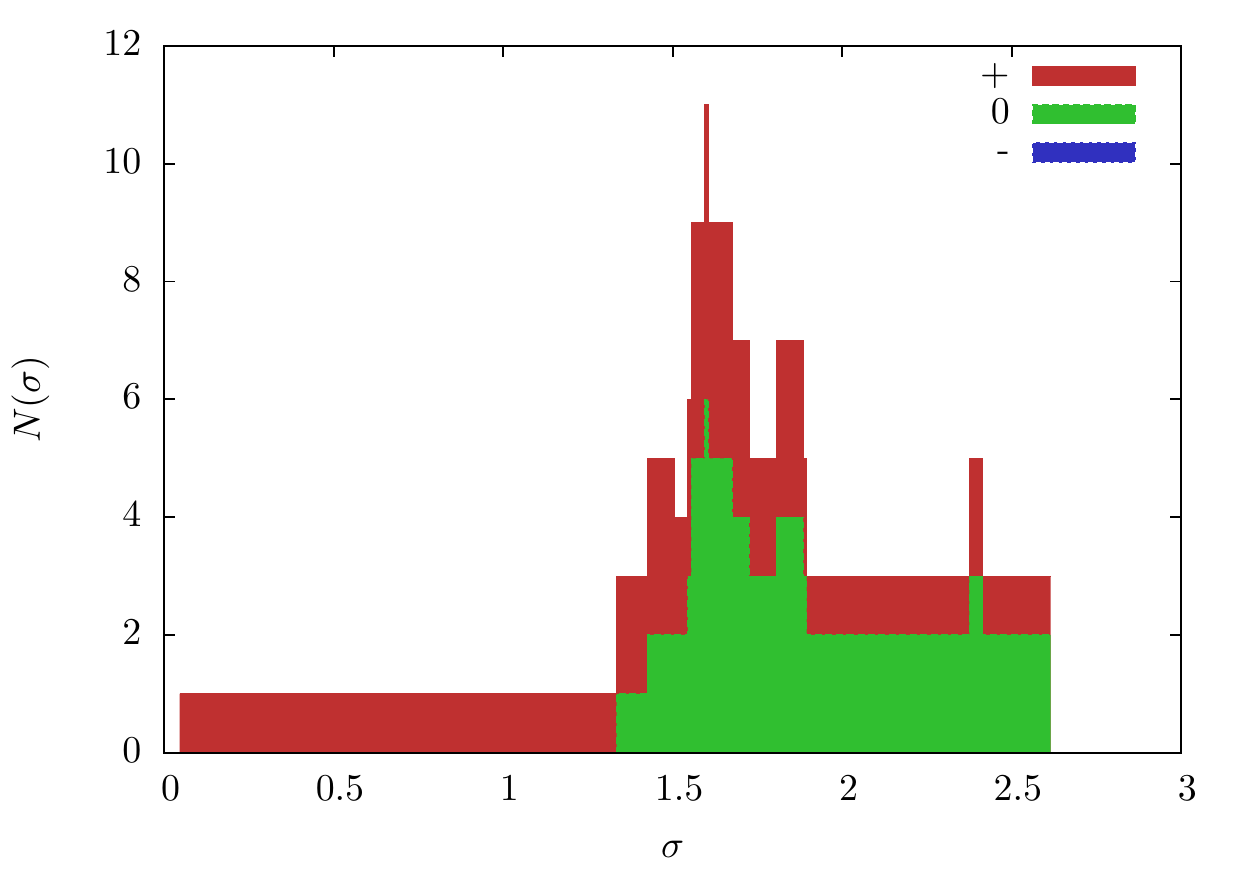}
  }
  \subfigure[Carreau-Yasuda TBD at $x_3$.]{
    \label{fig:tbd_cy3}
    \includegraphics[width=0.4\textwidth]{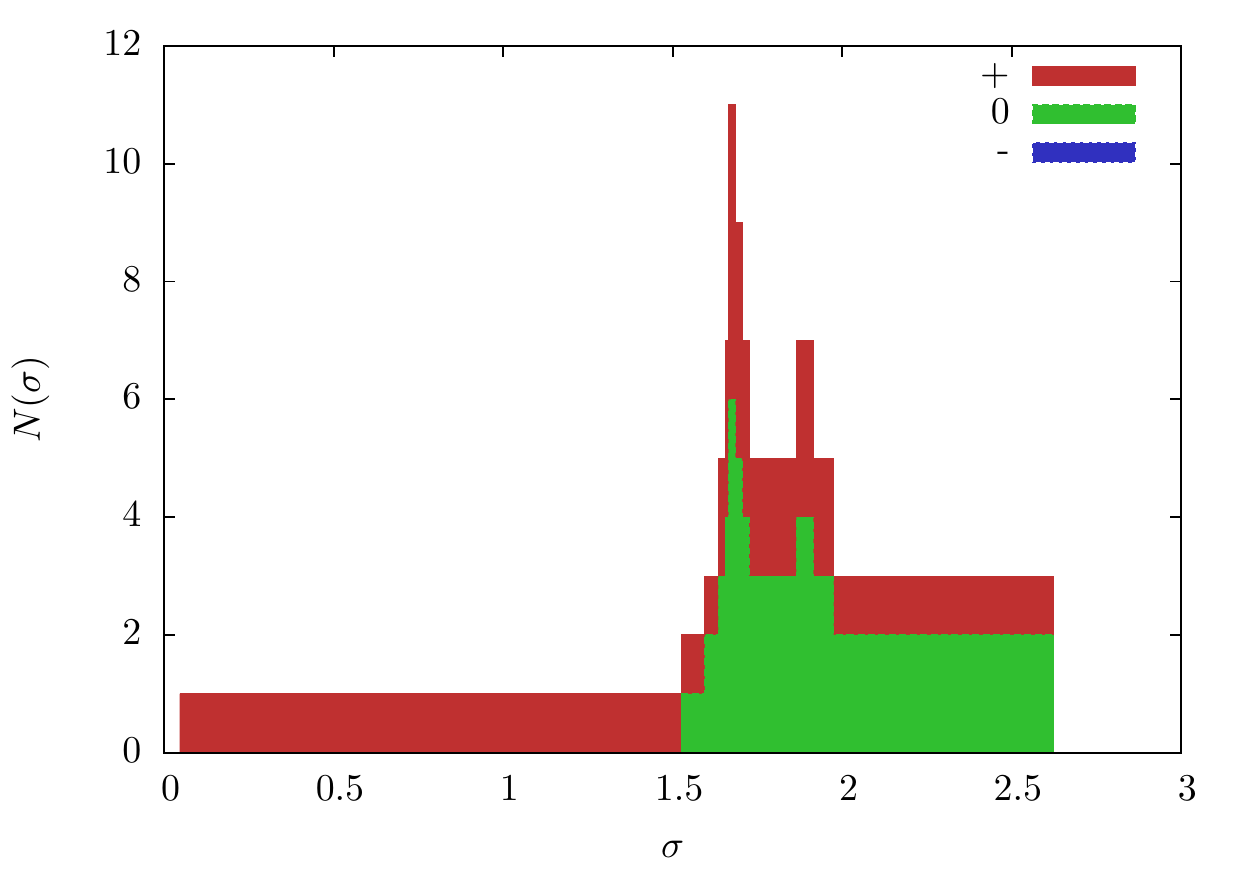}
  }
%   \caption{TBD analysis at point $x_3$.}
%   \label{fig:tbd3}
% \end{figure}
% %
% \begin{figure}
%   \centering
  \subfigure[Newtonian TBD at $x_4$.]{
    \label{fig:tbd_n4}
    \includegraphics[width=0.4\textwidth]{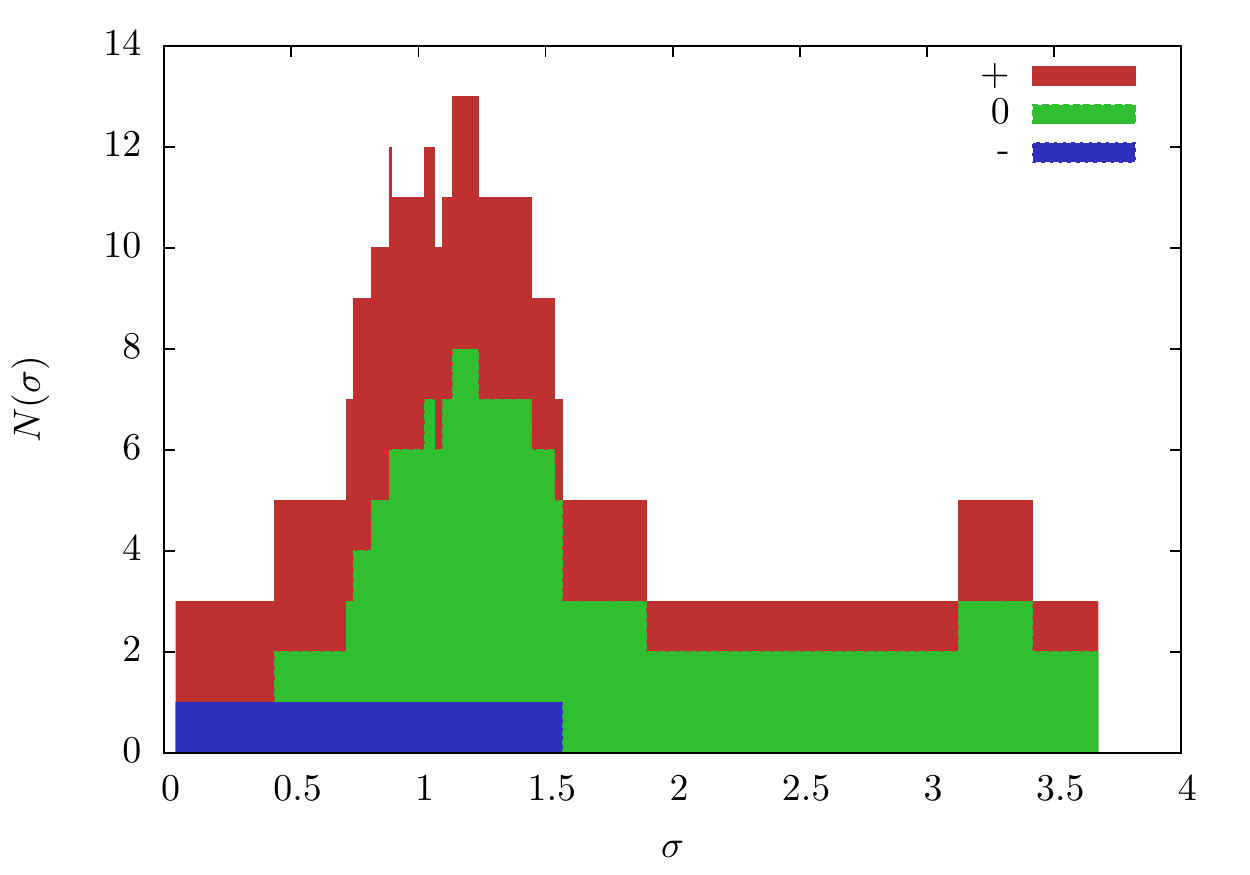}
  }
  \subfigure[Carreau-Yasuda TBD at $x_4$.]{
    \label{fig:tbd_cy4}
    \includegraphics[width=0.4\textwidth]{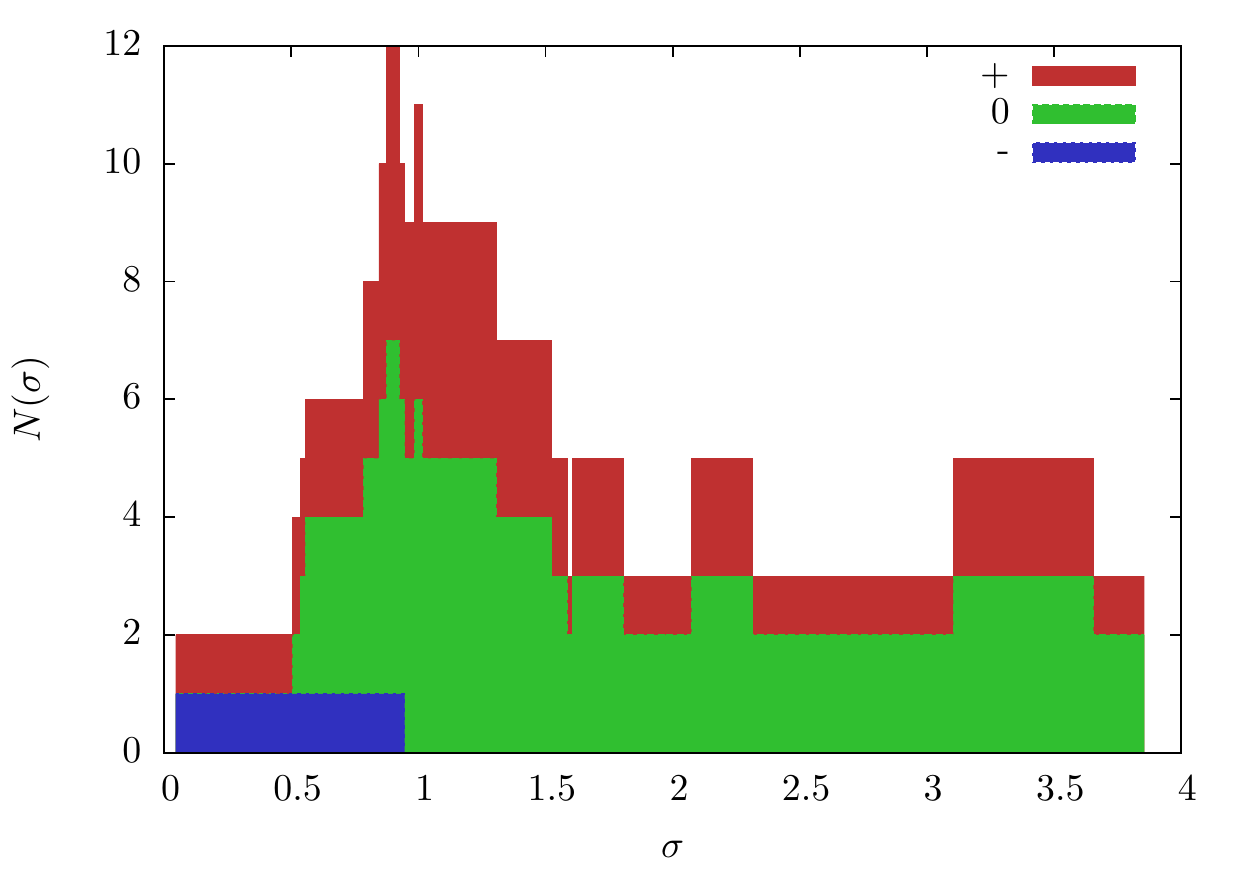}
  }
%   \caption{TBD analysis at point $x_4$.}
%   \label{fig:tbd4}
% \end{figure}
% %
% \begin{figure}
%   \centering
  \subfigure[Newtonian TBD at $x_5$.]{
    \label{fig:tbd_n5}
    \includegraphics[width=0.4\textwidth]{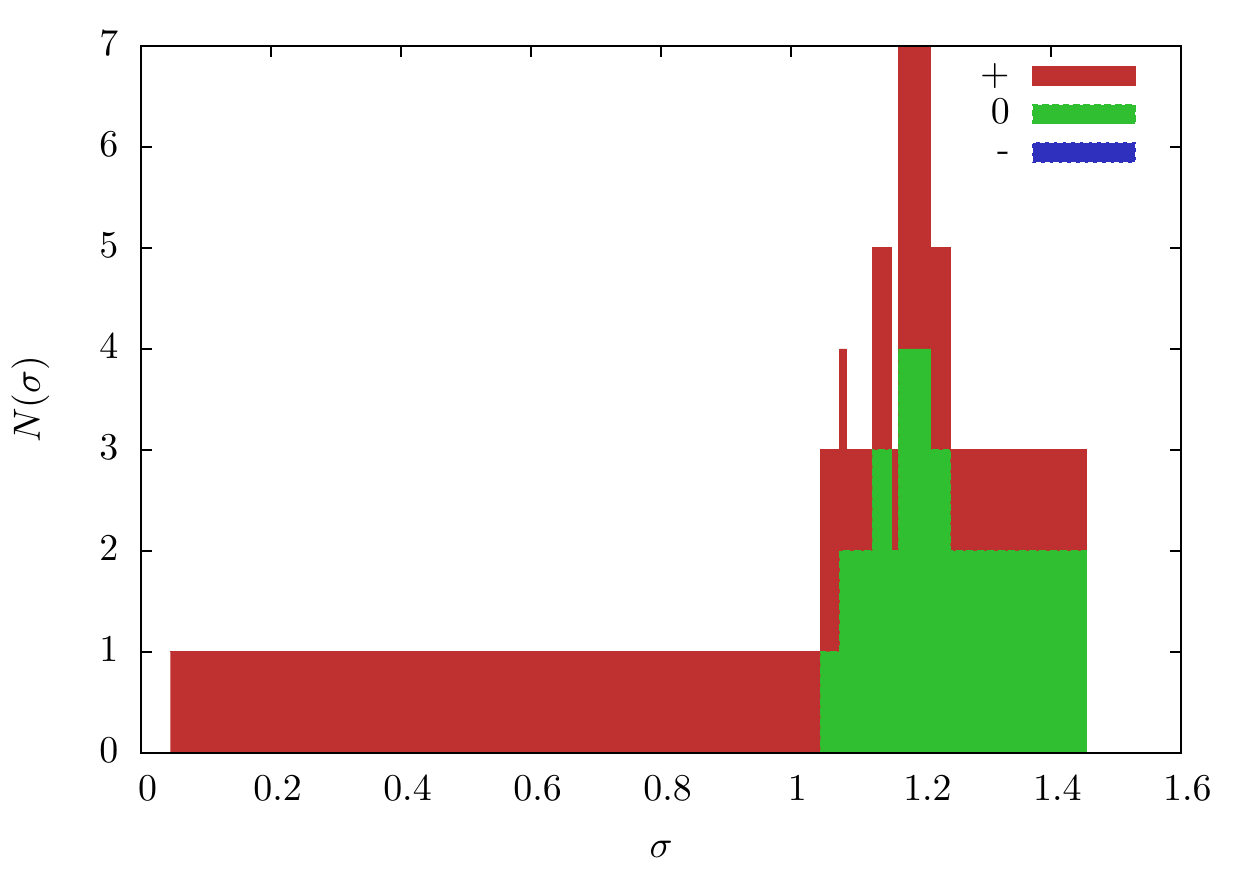}
  }
  \subfigure[Carreau-Yasuda TBD at $x_5$.]{
    \label{fig:tbd_cy5}
    \includegraphics[width=0.4\textwidth]{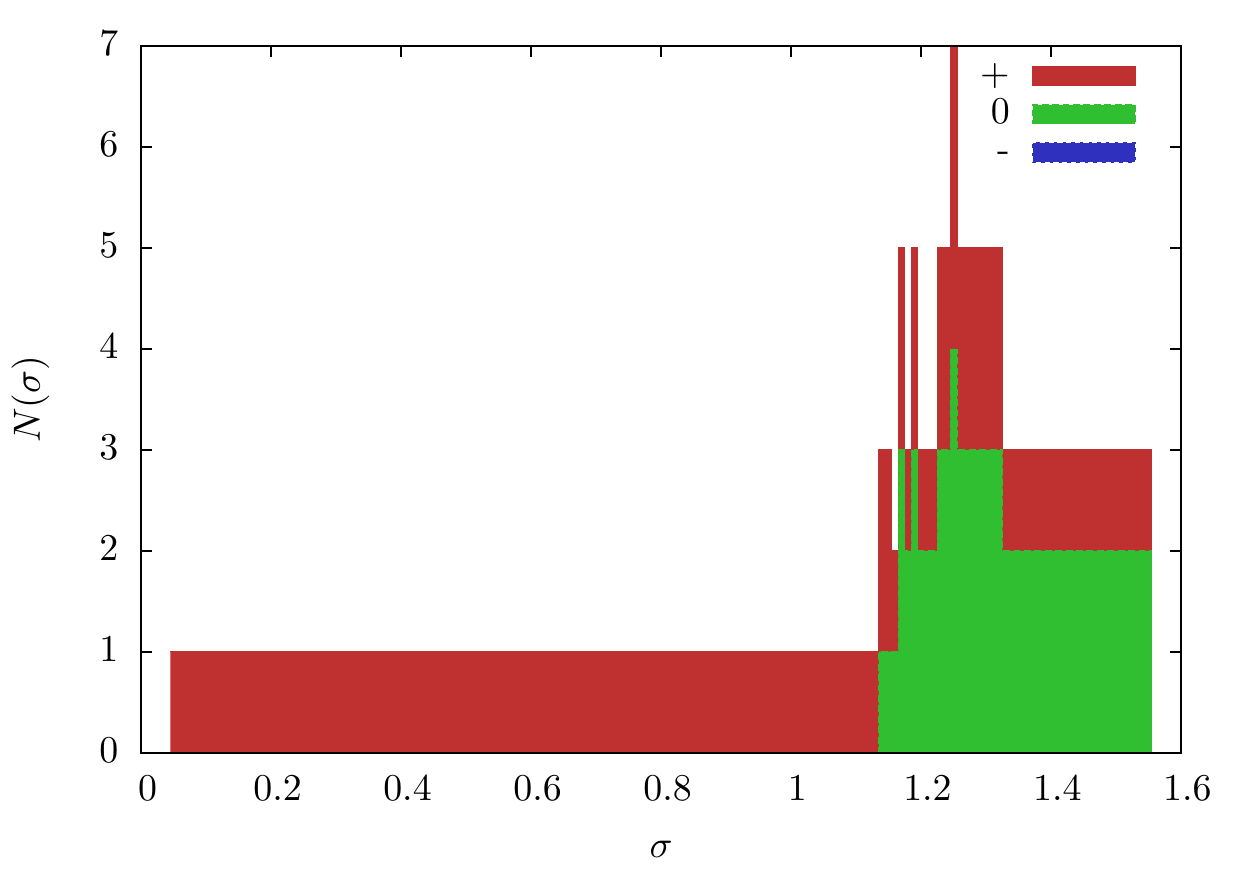}
  }
  \caption{TBD analysis at points $x_3$, $x_4$, and $x_5$. The values
    of $N^{\{-,0,+\}}(\sigma)$ are presented as a stacked histogram.}
  \label{fig:tbds}
\end{figure}

Figure \ref{fig:tbds} presents the TBD analysis at points $x_3$,
$x_4$, and $x_5$. The analysis confirms the change of sign of the WSS
signal (due to a change of direction in the traction field) occurring
at $x_4$ throughout the cardiac cycle (note the presence of a negative
component in figures \ref{fig:tbd_n4} and
\ref{fig:tbd_cy4}). Furthermore, the negative component covers a wider
range of threshold values in the Newtonian case. According to the
criteria outlined in \cite{gizzi11}, an oscillatory WSS signal is
considered healthy when the TBD analysis displays all three components
above a given critical threshold. In this case we observe a variation
of just over \SI{0.5}{\pascal} between the predicted largest healthy
threshold in the Newtonian and CY cases.

In the case of $x_3$ and $x_5$ no negative component appears in the
analysis indicating that there is no oscillation in the WSS
signal. This rules out one of the main factors correlated with vascular
disorders: oscillatory flow \cite{ku85}. Nevertheless, such signals
could still be considered risky if their mean value was below a given
threshold value \cite{shojima04}. We can clearly see in figures
\ref{fig:tbd_n3}--\ref{fig:tbd_cy3} and
\ref{fig:tbd_n5}--\ref{fig:tbd_cy5} that the choice of rheology model
would not have a significant impact on the assessment of risk since
both diagrams are nearly identical.

The points listed in table \ref{ta:sampling_points} which have not
been analysed in figure \ref{fig:tbds} fall within one of the two
previous scenarios.

\section{Conclusions}

In this work, we present a complete workflow for the simulation of blood
flow in a patient-specific 3D model of the right middle cerebral
artery. The main features of the model are: i) the geometry is
reconstructed from rotational angiography scans and discretised at
high resolution ($\Delta x=\SI{3.5e-5}{\metre}$), ii) inlet and outlet
boundary conditions are obtained with a 1D model of the complete
vascular system, and iii) the rheological properties of the blood can
be described with both Newtonian and generalized Newtonian
models. Simulations run efficiently on the HECToR supercomputer taking
\SIrange{53}{83}{\minute} for a domain comprising of 4,161,046 lattice
sites.

The workflow is applied to the comparison of two blood
rheology models: a Newtonian model ($\eta = \SI{3.5e-3}{\pascal\second}$) and
the Carreau-Yasuda (CY) model. This is done in a quantitative manner
in conjunction with
the recently proposed three-band diagram (TBD) analysis framework. 
% Our results show that
% despite the variations in haemodynamics extensively reported in the
% literature \cite{addcite}, when the Newtonian and the CY WSS signals obtained
% at a given spatial location are compared against a set of risk factors
% the differences are small. 
In agreement with previous work (see \emph{e.g.} \cite{morbiducci11, bernsdorf09}), our results show variations
in the haemodynamics recovered with each of the rheology models
studied. However, the evaluation of those results against a set of
risk factors with TBD show little to no difference.
In particular, a wall shear stress signal with strong
oscillatory component was found to be risky for thresholds equal to or greater
than 1.56 and \SI{0.94}{\pascal} for the Newtonian and CY models
respectively. In the case of non-oscillatory signals the analysis
returns almost identical results in both cases.

The main limitations of our study are: i) we have used a single geometry;
in order to achieve statistical significance a larger number of other cases
must be considered, and ii) the results presented may not hold in the
presence of intracranial aneurysms (ICAs) or other vascular
malformations where regions of much smaller shear rate may occur.
% , and
% iii) the complex relationship between haemodynamics and biology is
% neglected as the substrate for the initiation and evolution of
% vascular pathologies.

Our future lines of work include applying the analysis methodology
presented in the current work to a set of ICAs reconstructed from
rotational angiography images. We also plan to investigate the
coupling of the current 1D-3D haemodynamics model with agent-based
models of tissue remodelling using the multiscale modelling framework
presented in \cite{groen12b}.

\section*{Acknowledgements}
This work made use of HECToR, the UK's national high-performance
computing service, funded by the Office of Science and Technology
through EPSRC's High End Computing Programme. We also thank Dr Simone
Manini for support using the pyNS software.  This work was supported
by the British Heart Foundation, EPSRC grants ``Large Scale Lattice
Boltzmann for Biocolloidal Systems'' (EP/I034602/1) and 2020 Science
(http://www.2020science.net/, EP/I017909/1), and the EC-FP7 projects
CRESTA (http://www.cresta-project.eu/, grant no. 287703) and MAPPER
(http://www.mapper-project.eu/, grant no. 261507).

%\bibliographystyle{apalike} 
%\bibliography{main_biblio}

\end{document}